\documentclass[12pt,square]{iopart}
\usepackage{iopams}
\usepackage[dvips]{graphicx}
\usepackage{amssymb}
\usepackage{color}

\begin{document}

\title[Point-contact spectroscopy in Co-doped CaFe$_2$As$_2$]{Point-contact spectroscopy in Co-doped CaFe$_2$As$_2$:
nodal superconductivity and topological Fermi surface transition.}
\author{R S Gonnelli$^1$, M Tortello$^1$, D Daghero$^1$, R K Kremer$^2$, Z Bukowski$^{3,4}$, N D Zhigadlo$^3$, and J Karpinski$^3$}
\address{$^1$ Dipartimento di Scienza Applicata e Tecnologia, Politecnico di Torino,
corso Duca degli Abruzzi 24, 10129 Torino (TO) Italy}
\address{$^2$ Max-Planck-Institut für Festk\"{o}rperforschung, D-70569 Stuttgart (Germany)}
\address{$^3$ Laboratory for Solid State Physics, ETHZ, CH-8093 Zurich, Switzerland}
\address{$^4$ Institute of Low Temperature and Structure Research, Polish Academy of
Sciences, P.O. Box 1410, 50-422 Wroclaw, Poland}
\ead{renato.gonnelli@polito.it}

\begin{abstract}
We performed point-contact Andreev reflection spectroscopy
measurements in Ca(Fe$_{1-x}$Co$_x$)$_2$As$_2$ single crystals with
effective $x=0.060 \pm 0.005$.  The spectra of $ab$-plane contacts
show a zero-bias maximum and broad shoulders at about 5-6 meV. Their
fit with the three-dimensional Blonder-Tinkham-Klapwijk model
(making use of a analytical expression for the Fermi surface that
mimics the one calculated from first principles) shows that this
compound presents a large isotropic gap on the quasi-2D electronlike
Fermi surface sheets and a smaller anisotropic (possibly nodal) gap
on the 3D holelike Fermi surface pockets centered at the Z point in
the Brillouin zone. These results nicely fit into the theoretical
picture for the appearance of nodal superconductivity in 122
compounds.
\end{abstract}

\maketitle

\section{Introduction}
\label{sect:intro} Among the members of the 122 family of Fe-based
superconductors, the CaFe$_2$As$_2$ system is particularly
challenging from both the theoretical and the experimental points of
view. With respect to Sr-122 and Ba-122, which have been widely
studied in the past years and whose phase diagrams have been soon
determined, the Ca-122 system presents some peculiarities (namely, a
high sensitivity of the cell structure to chemical and physical
pressure, and a strong dependence of the physical properties of
single crystals on the growth procedure \cite{hu11}) that have
hindered its investigation and caused conflicts between data and
calculations by different groups.

Recently, the phase diagram of Ca(Fe$_{1-x}$Co$_x$)$_2$As$_2$ as a
function of $x$ has been determined independently by Hu et al.
\cite{hu11} and Harnagea et al. \cite{harnagea11}. Despite some
small quantitative disagreement on the Co content, the two phase
diagrams agree well with each other. In particular, unlike in the
sister compounds Sr(Fe$_{1-x}$Co$_x$)$_2$As$_2$ and
Ba(Fe$_{1-x}$Co$_x$)$_2$As$_2$, the superconducting phase appears
abruptly (at $x=0.03$ in Ref.\cite{hu11}) where the high-temperature
magnetic and structural transitions are still present, and the
critical temperature gradually decreases with increasing $x$.

In this paper we present the results of point-contact
Andreev-reflection spectroscopy (PCARS) measurements in
Ca(Fe$_{1-x}$Co$_x$)$_2$As$_2$ single crystals with $x=0.060 \pm
0.005$, aimed to determine the number, the amplitude and the
symmetry of the superconducting order parameter(s) in this compound.
These pieces of information, which were lacking up to now, are of
particular interest if compared to the predictions about the
dependence of the gap symmetry of 122 compounds on the shape of the
Fermi surface (and thus on the height of the pnictogen above the Fe
layer, $h_{As}$) \cite{suzuki11}. We will show indeed that the PCARS
results give evidence of a large isotropic gap and a small nodal (or
fully anisotropic) gap, and that DFT calculations of the Fermi
surface show a topological transition in the holelike Fermi surface
sheets occurring at the same doping content. We speculate that the
two facts may be correlated and that the symmetry of the OP on the
closed holelike pockets at $x\geq 0.06$ can be modeled as a 3D
$d$-wave one, with two nodal planes intersecting each other along
the $\Gamma$-Z line.

\section{Experimental results}\label{sect:experimental}
\begin{figure}
\begin{center}
\includegraphics[width=0.6\columnwidth]{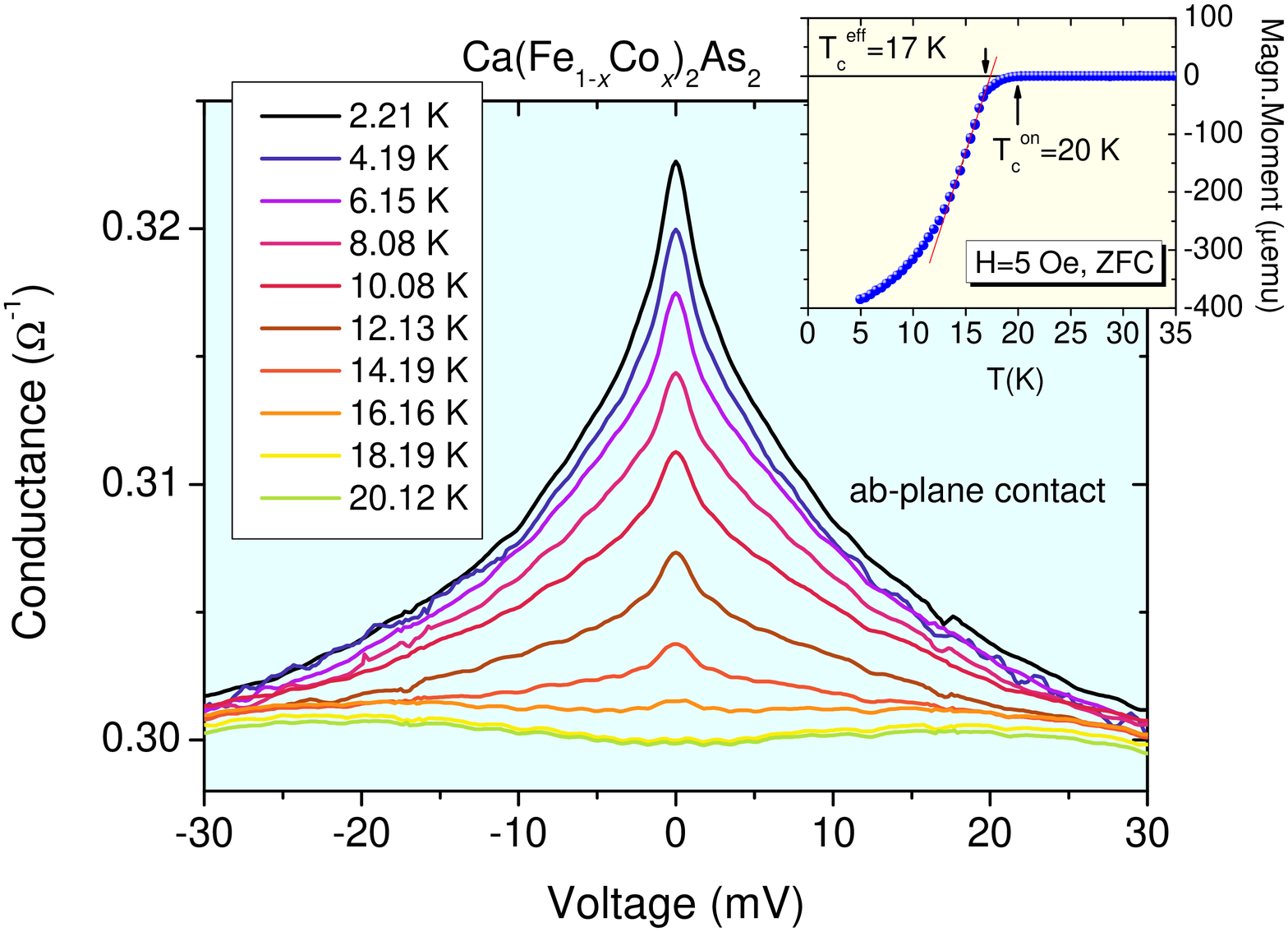}
\end{center}
\caption{Some experimental conductance curves (not normalized) as a
function of temperature. The point-contact resistance is about 3
$\Omega$. Inset: the zero-field-cooled DC magnetization of one of
our crystals.}\label{fig:1}
\end{figure}

The  Ca(Fe$_{1-x}$Co$_x$)$_2$As$_2$ single crystals used in this
work were grown from Sn flux, as described in Ref.
\cite{matusiak10}; they looked very similar to those shown in Ref.
\cite{harnagea11}, i.e. they were plate-like, with mirror surfaces
and the $c$ axis perpendicular to the plate. The inset to
Fig.\ref{fig:1} shows the DC magnetization of one of the crystals as
a function of temperature. The transition sets in at $T_c^{on}=20$ K
and an effective $T_c^{eff}=17$ K can be determined by extrapolating
the linear part of the curve. The rather broad transition is common
to all the state-of-the-art single crystals grown at present
\cite{hu11,harnagea11,abdel11,kumar09}. The broadening might be
related to chemical inhomogeneities over nanoscopic length scales
\cite{harnagea11} that are not detectable by energy- or
wavelength-dispersive X-ray spectroscopies \cite{harnagea11,hu11},
or might be intrinsic to the Ca-122 system because of its high
sensitivity to chemical and physical pressure \cite{harnagea11}. In
any case, the large transition width is not detrimental to
point-contact measurements since PCARS is a local probe not only of
the superconducting order parameter but also of the critical
temperature. The critical temperature of the best contacts always
fell between $T_c^{eff}$ and $T_c^{on}$. By comparison with the
phase diagram by Hu \emph{et al.} \cite{hu11}, we can conclude that
the real Co content of our crystals is $x=0.060 \pm 0.005$.

The point-contact measurements were performed by using the ``soft''
technique described elsewhere, i.e. by putting a small drop of Ag
paste ($\phi \simeq 50 \mu$m) on a fresh side surface of the crystal
(exposed by breaking it) so that the current is mainly injected
along the $ab$ planes. Contacts made in this way are more
mechanically and thermally stable than those made by using the
standard needle-anvil technique (i.e. by gently pressing a sharp
metallic tip against the same surface). The $I-V$ characteristics of
the normal metal/superconductor junctions obtained in this way were
measured and numerically differentiated to obtain the conductance
curves, $\mathrm{d}I/\mathrm{d}V$ vs. $V$.

Figure \ref{fig:1} shows the temperature dependence, from 2.21 K up
to 20.12 K, of the raw conductance curves of a $ab$-plane point
contact whose normal-state resistance was $R_N \simeq 3$ $\Omega$.
The Andreev signal disappears completely at $T_c^A \simeq 18$ K,
leaving a slightly bent but rather symmetric normal-state
conductance. Despite the rather low resistance, the conduction
through the contact is very likely to be ballistic. This is
witnessed by the fact that the high-energy tails of the curves
coincide with one another (which means that there is no contribution
from the Maxwell term in the contact resistance \cite{daghero10}),
by the absence of features associated with the current-induced
suppression of superconductivity (such as characteristic dips
\cite{sheet04}) and by the fact that the Andreev critical
temperature $T_c^A$ is not suppressed with respect to the bulk
$T_c$. The small resistance of the contact can indeed be reconciled
with the evidence of ballistic conduction if multiple parallel
nanocontacts are formed between Ag grains in the paste and the
crystal surface, which is not only very reasonable but practically
unavoidable in our point contacts.

Even at a first glance, the shape of the curves in Fig.\ref{fig:1}
looks not to be compatible with a nodeless gap (or multiple nodeless
gaps, as observed in most 1111 compounds \cite{daghero11} and in 122
compounds like Ba(Fe,Co)$_2$As$_2$ \cite{tortello10} and
(Ba,K)Fe$_2$As$_2$ \cite{szabo09}). It must be noted that this shape
is common to all the conductance spectra in the Andreev-reflection
regime we obtained in different $ab$-plane contacts, so that it
appears to be an intrinsic feature of the material and not an
artifact. A narrow maximum at zero bias such as that shown in the
curves in Fig.\ref{fig:1} has been observed in many unconventional
superconductors, including iron pnictides
\cite{lu10,yates08a,samuely09a}; however, this kind of feature can
have either intrinsic or extrinsic origins. For example, it can be
the hallmark of a Josephson supercurrent -- but this requires either
the formation of some SIS junction, or intrinsic Josephson effects
as in $c$-axis tunnelling in layered materials, both excluded here.
Alternatively, the zero-bias peak can also be due to the
non-perfectly ballistic conduction (already excluded for our
contacts), or the local damage or deformation of the surface due to
the pressure applied by the tip (see \cite{daghero10} and references
therein), but this is clearly impossible with our technique. Finally
-- and we claim this is the case in our measurements -- it can be a
direct consequence of the anisotropy of the superconducting order
parameter.

If the OP has nodal lines crossing the Fermi surface, as in the
$d$-wave or in the nodal $s$ symmetries proposed for 1111 pnictides
in suitable conditions \cite{kuroki09}, its change of sign can give
rise -- for suitable directions of current injection -- to
constructive interference between hole-like and electron-like
quasiparticles that result in zero-energy bound states localized at
the interface \cite{kashiwaya00}. These, in turn, manifest
themselves in a zero-bias peak in the Andreev-reflection
conductance. In principle, also the 3D nodal symmetry of the OP
proposed for the isovalent-doped Ba-122 system
\cite{suzuki11,graser10} could give rise to these effects. A similar
peak can also appear in diffusive metal/nodal superconductor
junctions as a result of the coherent Andreev reflection in the
former \cite{tanaka04}.

We have however recently shown \cite{daghero11} that, in the
presence of a large spectral broadening (as in our contacts on
$\mathrm{Ca(Fe,Co)_2As_2}$) it becomes impossible to distinguish a
true zero-bias conductance peak (due to a nodal OP) from the
zero-bias maximum that is the natural consequence of a OP with
zeroes on the Fermi surface. Strictly speaking, we can thus conclude
that the shape of the conductance curves of Fig.\ref{fig:1}
indicates that at least some component of the OP is zero somewhere
on the Fermi surface -- although we cannot say whether it changes
sign or not.


\begin{figure}
\begin{center}
\includegraphics[width=0.6\columnwidth]{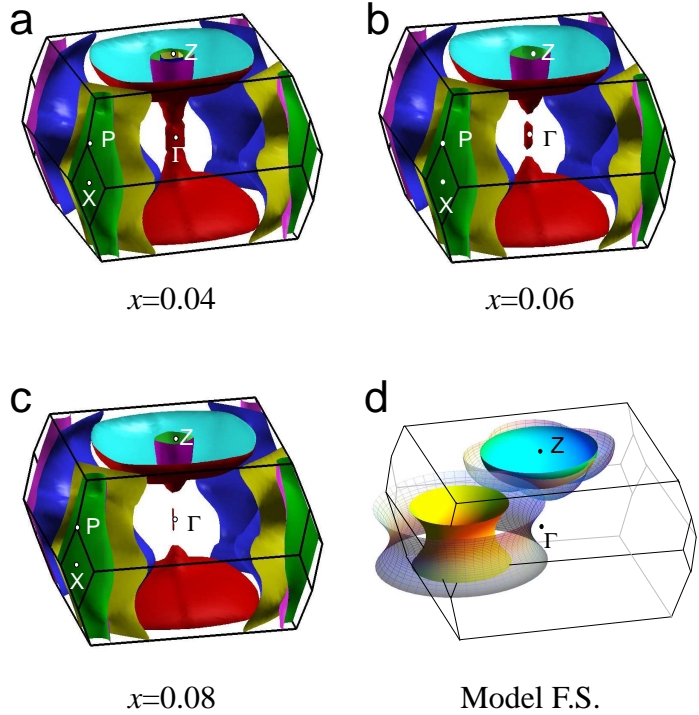}
\end{center}
\caption{(a) The Fermi surface of
Ca(Fe$_{0.96}$Co$_{0.04}$)$_2$As$_2$ calculated as described in the
text (Elk FP-LAPW code, GGA + VCA approach) by using the
low-temperature lattice parameters of the collapsed tetragonal
structure of CaFe$_2$As$_2$ from Ref.\cite{colonna11}. (b, c) the
same as in (a) but for $x=0.060$ and $x=0.080$, respectively. (d)
Model Fermi surface used to perform the 3D-BTK fit of the
experimental PCAR curves (matt surfaces) and the relevant amplitude
of the order parameter (gridded surfaces). Only one half of the two
Fermi surfaces is shown.}\label{fig:2}
\end{figure}

To extract quantitative information on the number, amplitude and
symmetry of the OPs (with the aforementioned limitations), it is
necessary first to normalize the experimental conductance curves and
then to compare them with the predictions of suitable models for
Andreev reflection. The first step is straightforward in this case,
since the normal-state conductance is well defined and, judging from
the superposition of the high-energy tails of the curves, seems not
to depend very much on temperature below $T_c^A$. Therefore, we will
divide every conductance curve at $T<T_c^A$ by a curve measured just
above $T_c^A$. In the case of Fig.\ref{fig:1}, we used the curve at
18.19 K. As for the fit of the normalized conductance curves, one
might choose for simplicity a two-band 2D-BTK model
\cite{kashiwaya00}. This model is actually based on the assumption
that the Fermi surface is approximately cylindrical, but in 122
compounds a more or less pronounced warping of the Fermi cylinders
is present already in the parent compounds, and increases with
doping \cite{graser10}. In the next section we will show that the 2D
approximation for the Fermi surface is not justified in
$\mathrm{Ca(Fe_{0.94}Co_{0.06})_2As_2}$ so that the more complicated
two-band 3D-BTK model must be used instead.

\section{Calculation of the Fermi surface}
To gain further insight into the real band structure of
Ca(Fe$_{1-x}$Co$_x$)$_2$As$_2$ we performed first-principle DFT
calculations by using the Elk FP-LAPW Code
($\mathrm{http://elk.sourceforge.net/}$) within the GGA approach for
the exchange correlation potential.  A virtual-crystal approximation
(VCA) was used to mimic the Co doping, owing to the fact that VCA
works particularly well for Fe-Co substitutions. In the absence of
direct experimental information on the low-temperature lattice
constants of Co-doped CaFe$_2$As$_2$, we used the values $a=b=3.925$
{\AA} and $c=11.356$ {\AA}, calculated in Ref. \cite{colonna11}
\emph{at low temperature}  for the collapsed tetragonal phase of
CaFe$_2$As$_2$ and obtained from the total energy minimization by
considering the stripe antiferromagnetic order. Owing to the small
doping dependence of the lattice constant at high temperature
\cite{hu11}, we assumed these values to be a good first
approximation to the real ones at the doping content of our interest
($x<0.08$). The values of the theoretical parameters are in good
agreement with the experimental ones measured in the collapsed
tetragonal phase of the parent compound at 40 K (and pressure
$P=0.5$ GPa \cite{mittal11}). Starting from the calculated
equilibrium phase and always considering the antiferromagnetic
phase, an optimized internal parameter $z_{As}= 0.7306$ (such that
$h_{As}=1.309${\AA}) was obtained. The charge density was thus
integrated over $8 \times 8\times 4$ $k$ points in the Brillouin
zone and the band structures as well as the Fermi surfaces were
calculated in the non-magnetic body-centered tetragonal phase.

Figures \ref{fig:2}(a), (b) and (c) show the Fermi surface (plotted
on a $40 \times 40\times 40$ $k$-points grid) of
Ca(Fe$_{1-x}$Co$_x$)$_2$As$_2$ for $x=0.04$, $x=0.06$ and $x=0.08$,
respectively. The evolution of the holelike Fermi surface towards a
full 3D nature on increasing $x$ is clear. In the $x=0.04$ case
(panel a), three holelike FS sheets are present: two inner closed
pockets centered at Z and an outer, strongly warped cylinder
extending along the $\Gamma$-Z line. At $x=0.06$ the smaller pocket
disappears and the warped cylinder undergoes a topological
transition splitting into separated closed surfaces centered at Z.
It is worth noting that this trend is qualitatively robust against
the details of the calculations, although some refinement (e.g. the
relaxation of the cell parameters or their small doping dependence)
could slightly change the doping content at which the complete
separation of the two closed sheets around Z occurs.

\section{Analysis of the experimental data}
In these conditions, it is necessary to use the complete 3D-BTK
model \cite{daghero11} to fit our conductance curves. We first
schematized the actual Fermi surface of
$\mathrm{Ca(Fe_{0.94}Co_{0.06})_2As_2}$ by means of a one-sheeted
hyperboloid of revolution (that simulates the electronlike FS sheets
centered in X) and an oblate spheroid (to simulate the 3D holelike
FS centered in Z). Fig.\ref{fig:2}(d) shows one half of the two
model Fermi surfaces (corresponding to the upper half of the real
Brillouin zone shown in panel (b)). The proportions between the size
of the two FSs have been respected, though the distance between them
(which is irrelevant in the following calculation) has been
increased for clarity with respect to the actual distance shown in
fig.\ref{fig:2}b. For this reason the Brillouin zone sketched in
Fig.\ref{fig:2}d appears wider that the actual one and should be
just considered as a guide to the eye.

Before proceeding, we had to make a guess about the symmetry of the
two OPs residing on the two FSs. According to recent theoretical
predictions, in the Ba-122 system an evolution from a pure $s\pm$
gap symmetry towards a peculiar 3D nodal one is expected when the
pnictogen height $z_{As}$ is reduced, for example by isovalent P
doping \cite{suzuki11}. It is interesting to note that the reduction
in $z_{As}$ simply leads to a larger holelike Fermi surface around
the Z point, similar to the one depicted in Fig. \ref{fig:2}(a). In
these conditions, a three-dimensional sign change of the OP takes
place within this Fermi surface near the edge of the Brillouin zone
\cite{suzuki11,graser10}, while the gap on the electron Fermi
surface always remains fully open. When the holelike FS is further
deformed into separated closed surfaces, as in our case, the
symmetry of the OP relevant to it must evolve as well. We guess that
the result of this evolution can be expressed as a 3D $d$-wave OP,
whose expression in the reciprocal space is $\Delta(\theta,
\phi)=\Delta \cos(2\theta)\sin^2(\phi)$, $\theta$ being the
azimuthal angle in the ($k_x, k_y$) plane of the reciprocal space
and $\phi$ the inclination angle. This OP features two orthogonal
nodal planes intersecting along the $\Gamma$-Z line. As for the
electronlike FS sheets, we assume that the relevant OP keeps a
$s$-wave symmetry. The amplitude of the two OPs are shown in
Fig.\ref{fig:2}(d) as gridded surfaces.


Now the theoretical conductance curve can be calculated, within some
reasonable assumptions, by means of equation 9 of
Ref.\cite{daghero11}. The two bands will be indicated by the
subscripts 1 (holelike band) and 2 (electronlike band). The degrees
of freedom of the problem are 7: the OP amplitudes $\Delta_1$ and
$\Delta_2$, the barrier parameters $Z_1$ and $Z_2$, the broadening
parameters $\Gamma_1$ and $\Gamma_2$, plus the angle $\alpha$
between the direction of current injection (here in the basal plane)
and the lobes of the $d$-wave gap $\Delta_1$. The relative weight of
the two bands in the conductance is not an adjustable parameter but
directly follows from the geometry of the problem (i.e. the shape of
the relevant Fermi surface sheets and the direction of current
injection) and from the choice of the $Z$ parameters.

\begin{figure}
\begin{center}
\includegraphics[width=0.6\columnwidth]{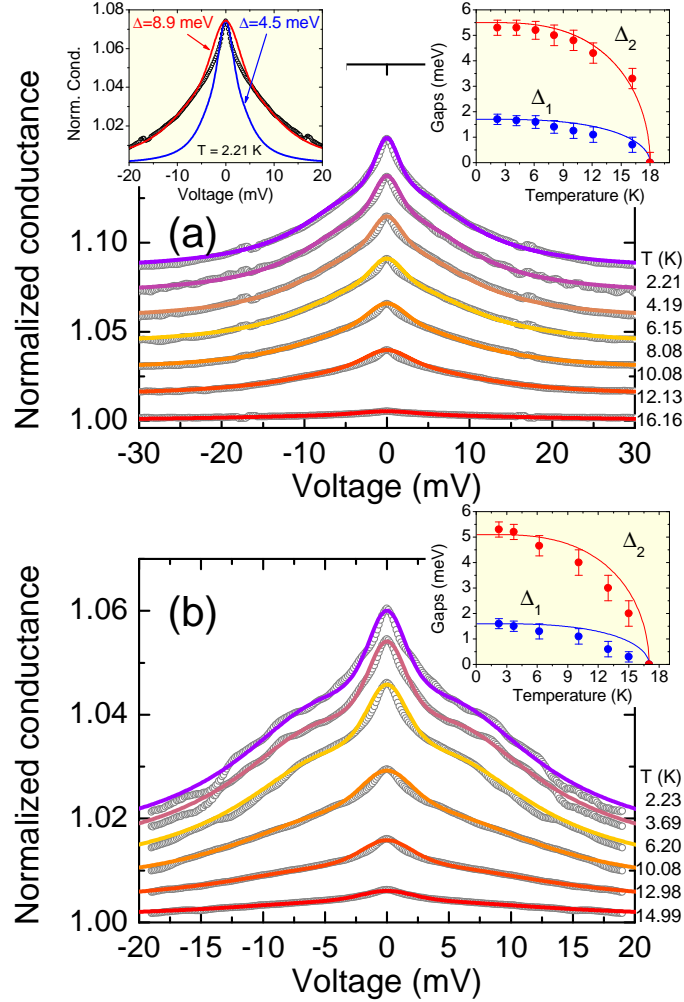}
\end{center}
\caption{(a) Normalized conductance curves of the same contact as in
Fig. \ref{fig:1} (symbols) and their relevant 3D-BTK fit obtained by
using the model Fermi surface of Fig.\ref{fig:2}d (lines). The left
inset shows two different one-gap, $d$-wave BTK fits (lines) to the
low-temperature curve of the main panel (symbols). The values of the
fitting parameters are: $\Delta=8.9$ meV, $\Gamma=6.2$ meV, $Z=1.1$,
$\alpha=\pi/8$ for the red, wider curve, and $\Delta=4.5$ meV,
$\Gamma=3.2$ meV, $Z=1.1$, $\alpha=\pi/8$ for the blue narrow curve.
(b) Same as in (a) but for a different contact, with normal-state
resistance $R_N \simeq 3.4 \Omega$. Both in (a) and (b) the right
insets report the values of the gap amplitudes $\Delta_1$ and
$\Delta_2$ (circles) as a function of temperature. Lines represent
two BCS-like temperature dependencies of the usual approximated form
$\Delta_i(T)=\Delta_i(0)\tanh(1.74\sqrt{(T_c / T)-1})$. }
\label{fig:3}
\end{figure}

Figure \ref{fig:3}(a) shows some of the experimental conductance
curves of Fig.\ref{fig:1}, normalized and compared to the
best-fitting curves (lines) calculated within the aforementioned
model. The parameters that allow fitting the low-temperature curve
are the following: $\Delta_1=1.7$ meV, $\Delta_2=5.3$ meV,
$Z_1=0.4$, $Z_2=0.185$, $\Gamma_1=1.66$ meV and $\Gamma_2=6.4$ meV.
The high values of the $\Gamma$ parameters are due to the smallness
of the Andreev signal. Such large values could be avoided by
introducing an additional scaling parameter S sometimes used in
these cases \cite{naidyuk11}, for instance to account for a fraction
of injected current that does not give rise to Andreev reflection.
To reproduce the ZBCP a misorientation angle $\alpha=\pi/8$ was
necessary. This value is quite reasonable because the side surfaces
of the crystals, on which the point contacts are made, are rather
irregular and thus the angle of current injection in the basal plane
is not well defined. Indeed, it can be shown that the conductance
curve one would obtain by averaging over all angles $\alpha$ between
$0$ and $\pi/4$ is best described by using a single $\alpha \simeq
\pi/8$. Note that the values of the gaps given by this fit are very
similar to those obtained by using a simplified 2D BTK model as in
Ref.\cite{daghero11}, but the $Z$ values are systematically smaller
because of the $Z$-enhancing effect discussed elsewhere
\cite{daghero11}. Moreover, as already explained, similar results
could be obtained by fitting the experimental normalized curves with
a $s$-wave OP on one band and a \emph{fully anisotropic} $s$-wave OP
(with zeros on the Fermi surface) on the other band. It is also
worth noting that a simpler model with a smaller number of
parameters fails to reproduce the experimental conductance curves.
The left inset to Fig.\ref{fig:3}(a) shows that a single-gap
$d$-wave model (with 4 parameters) can only reproduce either the
zero bias maximum (blue line) or the higher energy portion of the
conductance curve (red line). Though the latter curve is better, the
one-band fit is still less accurate than the two-band one; moreover,
it would give an unreasonably large value of the gap.

Fig.\ref{fig:3}(b) shows the temperature dependence of the
normalized conductance curves (symbols) of a different point
contact, whose normal-state resistance was $R_N \simeq 3.4\,
\Omega$. As in panel (a), the experimental data are compared to the
3D-BTK best-fitting curves calculated for a large isotropic gap and
a smaller $d$-wave gap. The fitting parameters at the lowest
temperature are $\Delta_1=1.6$~meV, $\Delta_2=5.3$~meV, $Z_1=0.35$,
$Z_2=0.265$,$\Gamma_1=2.2$~meV, $\Gamma_2=6.8$~meV and finally
$\alpha=\pi/7.5$. The values of the broadening parameters are even
higher than in the previous case, because the signal is smaller, but
the values of the gaps are in very good agreement with those
obtained from the curves in Fig.\ref{fig:3}(a). Some small
additional structures are also visible in the lowest-temperature
curves at about 8 and 12 meV that are not reproduced by the fit.
These features might be related to a strong electron-boson
interaction, as shown in Refs. \cite{daghero11,tortello10}. However,
in this case we did not observe them in a reproducible way, and
moreover they disappear soon with increasing temperature and the
Andreev signal is too small to draw any conclusion in this sense. It
is worth mentioning that such features contribute very little to the
uncertainty of the gap values, which mostly arises from the
smallness of the Andreev signal, as discussed below.

As shown in Fig.\ref{fig:3}(a) and (b), we were able to fit the
conductance curves at any temperature up to $T_c^A$ by using the
same model and keeping the barrier parameters ($Z_1$ and $Z_2$) and
the angle $\alpha$ independent of temperature. The amplitudes of the
gaps, $\Delta_1$ and $\Delta_2$, are shown as a function of
temperature in the right insets to Fig.\ref{fig:3}(a) and (b). Error
bars indicate the spread of gap values obtained in different fits of
the same curve, with different combinations of the remaining
parameters ($Z_i$, $\Gamma_i$ and $\alpha$). Because of the
smallness of the Andreev signal, this spread at low temperature is
of the order of 0.2 meV on $\Delta_1$ and 0.3 meV on $\Delta_2$, and
increases on heating. For the sake of comparison, the same insets
also show two BCS-like trends -- approximately expressed as
$\Delta_i(T)=\Delta_i(0)\tanh(1.74\sqrt{(T_c^A / T)-1})$, where
$\Delta_i(0)$ is the low-temperature gap and $T_c^A$ the
experimental critical temperature of the contact. While the data of
panel (a) look fairly compatible with these trends, some suppression
of the gaps are observed in the inset to panel (b) in the proximity
of $T_c^A$. The gap ratios turn out to be $2\Delta_1/k_B
T_c^A=2.2-2.3$ (much smaller than the BCS value for $d$-wave
superconductors) and $2\Delta_2/k_B T_c^A=6.85-7.25$, which is
instead suggestive of a strong electron-boson coupling.

\section{Conclusions}
We have presented the results of point-contact Andreev reflection
measurements in state-of-the-art Ca(Fe$_{1-x}$Co$_x$)$_2$As$_2$
single crystals with effective $x=0.060\pm 0.005$. The point
contacts were made on the side surface, i.e. the current was mainly
injected along the $ab$ planes. We have shown that the shape of the
PCARS spectra can be nicely explained within a multiband (multigap)
scenario where a large isotropic gap $\Delta_2$ coexists with a
smaller nodal (or fully anisotropic) gap $\Delta_1$. These two gaps
can be associated to the electronlike and holelike Fermi surface
sheets, respectively, on the basis of recent theoretical papers
\cite{graser10,suzuki11} that predict the development of 3D nodes in
the OP of the holelike FS when the size of the latter around the Z
points of the Brillouin zone increases (for example, in Ba-122, as a
consequence of P doping). We have shown by first-principle DFT
calculations that Co doping in CaFe$_2$As$_2$ has a similar, but
even more pronounced, effect on the holelike FS, which actually
undergoes a topological transition at about $x\simeq 0.06$ from a
warped cylinder along the $\Gamma$-Z line to separate ellipsoids
centered around the Z points. In these conditions, we have assumed
that the symmetry of the nodal gap $\Delta_1$ can be described by a
3D $d$-wave function, while $\Delta_2$ keeps a $s$-wave symmetry.
The fit of the experimental spectra with a two-band 3D-BTK model
using a model Fermi surface that simulates the calculated one gives
$\Delta_2=5.3 \pm 0.3$ meV and $\Delta_1=1.6\pm 0.2$ meV.

In conclusion, our results seem to confirm and extend the existing
predictions about the evolution of the gap symmetry as a function of
isovalent doping, and might suggest a connection between the
topological transition of the holelike FS and the appearance of
nodal points in the gap along the $\Gamma$-Z line. Further
point-contact measurements in other Fe-based compounds and with the
current injected in different directions could further substantiate
this picture.

\section*{Acknowledgments}
The work at Politecnico di Torino was supported by the PRIN Project
No. 2008XWLWF9-005. The work at ETHZ was supported by the NCCR
Project MaNEP. The authors thank very much G. Profeta for
suggestions and enlightening discussions. R.S.G. acknowledges R. K.
Kremer and the Max Planck Institute for Solid State Research in
Stuttgart where the measurements were carried out.
\\
\vspace{1cm}

\section*{References}
\providecommand{\newblock}{}

\end{document}